# Charging Wireless Sensor Networks with Mobile Charger and Infrastructure Pivot Cluster Heads

Dongsoo Har

Abstract

Wireless rechargeable sensor networks (WRSNs) consisting of sensor nodes with batteries have been at the forefront of sensing and communication technologies in the last few years. Sensor networks with different missions are being massively rolled out, particularly in the internet-of-things commercial market. To ensure sustainable operation of WRSNs, charging in a timely fashion is very important, since lack of energy of even a single sensor node could result in serious outcomes. With the large number of WRSNs existing and to be existed, energy-efficient charging schemes are becoming indispensable to workplaces that demand a proper level of operating cost. Selection of charging scheme depends on network parameters such as the distribution pattern of sensor nodes, the mobility of the charger, and the availability of the directional antenna. Among current charging techniques, radio frequency (RF) remote charging with a small transmit antenna is gaining interest when non-contact type charging is required for sensor nodes. RF charging is particularly useful when sensor nodes are distributed in the service area. To obtain higher charging efficiency with RF charging, optimal path planning for mobile chargers, and the beamforming technique, implemented by making use of a directional antenna, can be considered. In this article, we present a review of RF charging for WRSNs from the perspectives of charging by mobile charger, harvesting using sensor nodes, and energy trading between sensor nodes. The concept of a pivot cluster head is introduced and a novel RF charging scheme in two stages, consisting of charging pivot cluster heads by a mobile charger with a directional antenna and charging member sensor nodes by pivot cluster heads with directional antennae, is presented.

## I. INTRODUCTION

Dongsoo Har is with the Cho Chun Shik Graduate School of Green Transportation, KAIST, 291 Daehak-ro, Daejeon, Republic of Korea.
E-mail: dshar@kaist.ac.kr (Dongsoo Har).

The Internet of things (IoT) is creating a new paradigm in regards to the connectivity and intelligence of various types of sensor devices. Advancements in microchip fabrication technology have led to widespread adoption of sensor devices, such as radio frequency identification (RFID) tags, and extensive deployment of wireless sensor networks. Wireless sensor networks, which treat sensor devices as network nodes, serve as critical enablers of the IoT [1], carrying out missions of monitoring, localization, tracking of objects by seamless transfer of data in real time between sensor nodes [2]. Since sustainable operation of wireless sensor networks is a major concern, particularly with mission-critical sensor networks, the importance of charging the sensor nodes in a timely manner cannot be overemphasized. Devices operated with batteries usually have short lifetimes and most wireless sensor networks cease their operation when even a single or small number of sensor nodes become energy-depleted. Therefore, one of the primary objectives considered when designing a wireless rechargeable sensor network (WRSN) consisting of sensor nodes with batteries is to prolong the network lifetime without compromising the sensing performance [3]. From this viewpoint, it is necessary to develop a charging scheme for WRSNs that does not jeopardize the quality of service of the network operation. Charging the WRSNs involves several issues. Charging method can be contact type near-field charging or non-contact type far-field charging. Contact type near-field charging is typically based on magnetic induction and low frequency in the HF band. Charging efficiency of this method is very low when the sensor node is located more than a meter from the charger. Charger can be an infrastructure charger or a mobile charger. An infrastructure charger or immobile charger is not suitable for WRSNs deployed in large service areas where the sensor nodes are distributed. Charging efficiency with mobile chargers is higher because energy transfer to individual sensor nodes suffers less wave propagation loss with small distances between the charger and the sensor nodes. The configuration of the WRSN depends on the number of sensor nodes and the type of node localization. When many sensor nodes, on the order of hundred, are distributed in a large service area, charging efficiency significantly varies according to the specific charging scheme. Sensor nodes of WRSNs must have energy harvesting capability. Energy harvesting circuits have different configurations depending on the type of energy source. When that source is ambient radio frequency (RF) energy from a TV signal, cellular signal, or Wi-Fi signal, the amount of harvested energy is typically very low. On the other hand, when it is RF energy transferred by an intended charger, the amount of harvested energy is much larger. RF (energy) charging is gaining interest because of the use of small transmit and receive antennae for energy transmission and reception. In addition, the beamforming technique, which can provide high directional gain for antennae, can facilitate increased charging efficiency.

When a decent number of sensor nodes are involved in a WRSN deployed in a large service area, RF charging with a mobile charger requires substantial charging time for all the sensor nodes to reach the target energy level. Since this charging process is repeated regularly or irregularly according to application of the WRSN, an energy-efficient charging scheme is indispensable to maintaining a proper level of operating cost. In this article, an energy-efficient charging scheme is considered a charging scheme with reduced charging time for a given transmit power of the charger. In order to achieve this goal, optimized charging schemes for WRSNs aiming at finding optimal charging spots [4] and establishing optimal velocity control [5] have recently appeared in the literature. However, with a large number of sensor nodes to recharge, optimization for energy-efficient charging is accompanied by high computational complexity, which grows non-linearly with increased number of sensor nodes. Therefore, a sub-optimal charging scheme with reduced computational complexity might be more effective in practice. To this end, a novel energy-efficient charging method in two stages with a mobile charger and pivot cluster heads is presented here.

All the sensor nodes considered here are stationary or quasi-stationary sensor nodes. For energy-efficient charging, the sensor nodes are classified into clusters. Each cluster consists of a cluster head and member sensor nodes. When the cluster head has a directional antenna capable of steering beamforming for transmitting energy, it is a pivot cluster head here. Other than the pivot cluster heads, the sensor nodes have omni-directional antennae with 0dB antenna gain. The RF charging scheme in two stages is composed of charging for pivot cluster heads in the first stage with a mobile charger and charging for member sensor nodes by pivot cluster heads in the second stage. This charging scheme is hierarchical, in contrast to other flat charging schemes. The mobile charger is only concerned with the pivot cluster heads, so the computational complexity is substantially reduced. The pivot cluster heads are determined according to a simple clustering algorithm and it is highly likely that pivot cluster heads will be located at or around the center of locally grouped sensor nodes. Due to the natural wavefront spreading of electromagnetic waves, member sensor nodes other than the pivot cluster heads are also charged in the first stage with the mobile charger. When the first stage is complete, all the pivot cluster heads become overcharged, i.e., charged more than the target level set for the pivot cluster heads, and member sensor nodes are overcharged or partially charged. To complete the entire charging process for all the sensor nodes, energy transfer from pivot cluster heads to member sensor nodes takes place until all the member sensor nodes are charged over the target level set for the member sensor nodes. To achieve an energy-efficient charging process, the target energy level of the pivot cluster heads and the target energy level of the member sensor nodes are set at different values. Notably, antennae with directive gain are used for the mobile charger and also the pivot cluster heads when these entities transfer energy. For a comparison purpose, another charging scheme without costly directional antennae for cluster heads is also considered. Main differences of this charging scheme are the target energy level of the cluster heads set differently from that of the pivot cluster heads and energy trading between overcharged sensor nodes, including cluster heads, and undercharged sensor nodes.

This article is organized as follows. The review of energy harvesting and RF charging with a mobile charger is given in Section II. The process of energy-efficient charging in two stages with a mobile charger and pivot cluster heads is presented in Section III. The conclusion of this article is given in Section IV.

## II. Energy Harvesting and RF Charging with Mobile Charger

### A. Energy Harvesting with Different RF Sources

Wireless remote charging is seen as a powerful technology that can be used to power various types of sensor devices with capabilities of energy harvesting. Figure 1 illustrates the types of energy sources for energy harvesting. Solar energy can be captured by sensor nodes with small photovoltaic panels installed on or around them. Thermal energy is harvested by making use of thermoelectric converters converting heat energy into electric energy according to the Seebeck effect [6]. Harvesting vibrational energy is enabled by piezoelectric sensors utilized as transducers. However, these types of environmental energy are vulnerable to intermittency. Solar energy is accessible only during daytime and under certain weather conditions; thermal energy is excessively scarce for energy demanding applications; sources generating vibrational energy are rather limited. For these reasons, RF energy is emerging as the major source to power sensor devices. Harvesting RF energy can be practiced with a battery, since most components necessary for harvesting, such as antenna, rectifier, and amplifier, already exist in typical sensor devices implemented in the form of customized ASICs or FPGAs [7, 8]. Harvesting ambient RF energy can be classified into two different processes, according to the type of energy source, as follows:

- Anticipated sources: RF sources that are always present, e.g., TV signals and radio signals, but are not optimized nor specialized for the transmission of power to the sensor devices
- Dynamic energy sources: RF sources that are periodic and unstable, usually transmitting in short periods of time and requiring spectrum sensing by sensor devices for harvesting opportunities

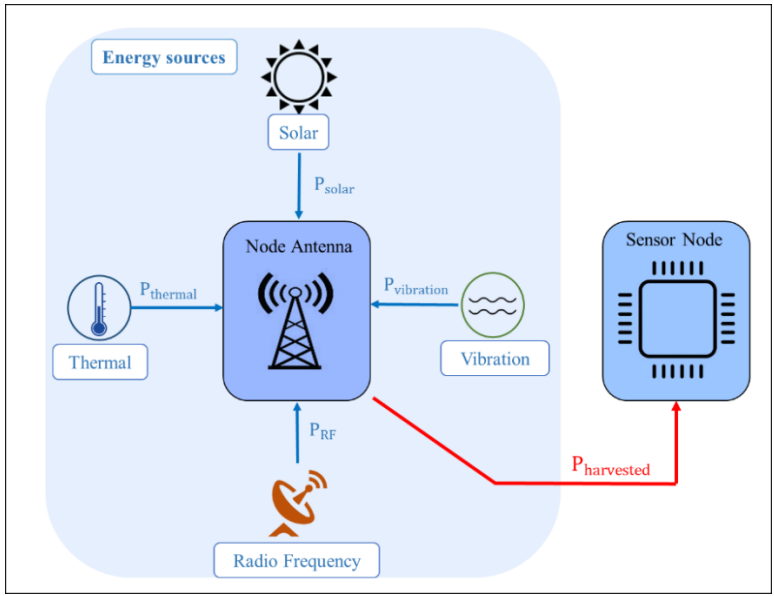

Fig. 1. Different types of sources for energy harvesting by sensor device.

On the other hand, harvesting RF energy with a dedicated RF charger is predictable with decent accuracy, when the pathloss model to estimate the power received at sensor devices and the location information of the sensor devices are commensurate with the desired accuracy. RF charging is expected to deliver a steady influx of energy to the sensor devices as long as the charger keeps transferring. It can be optimized in relation to charging frequency and the beamforming direction of the RF charger can be adjusted to obtain the highest level of charging efficiency. RF charging can work with tens of meters of distances between the transmitter and the sensor devices. Thus, RF charging can be applicable to operate the sensor nodes in WRSNs. An example of a sensor device capable of RF energy harvesting is shown in Fig. 2. The sensor device is seen to have a T-shaped directional antenna.

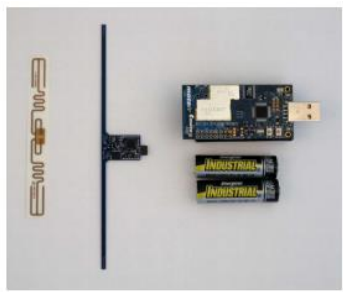

Fig. 2. Sensor node capable of RF energy harvesting [9].

## B. Energy Harvesting in Spectrum Sharing and Energy Trading

Signal relaying is another form of charging that is combined with harvesting. It can be used to transfer energy from one node to another. Part of the energy of the signal received by the relay node, denoted by $P_{rf}$ in Fig. 3, is used for harvesting and the remaining is used for information relaying. This process usually occurs either in a time switched or a power split configuration. The time switched configuration in three phases is shown in Fig. 3(a). With $\theta T$ being the time taken for energy harvesting, $(1-\theta)T/2$ in the second phase is for energy receiving and another $(1-\theta)T/2$ in the third phase is for data transmission. A power split configuration in two phases is shown in Fig.3(b). With $T/2$ being the time taken for both harvesting and receiving information, another $T/2$ in the second phase is for relaying the signal. Figure 4 shows an experimental setup of the signal relaying in Fig. 3 with a source node (RF source), a relay node (intermediate node), and a receiver node (RF analyzer).

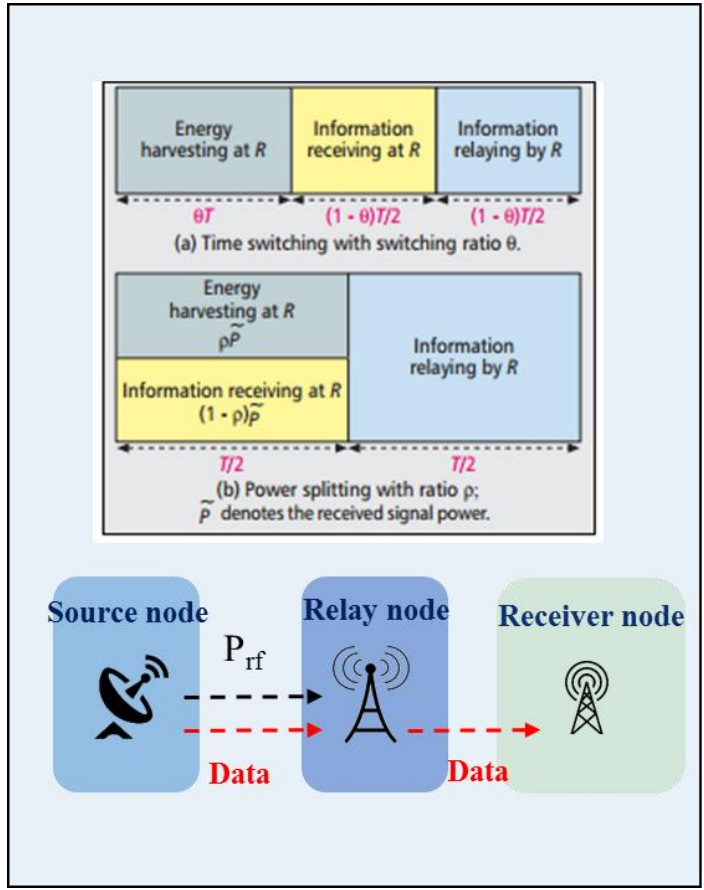

Fig. 3. Signal relaying combined with energy transfer [10]: (a) time switched energy transfer; (b) power split energy transfer. "R" indicates the relaying node.

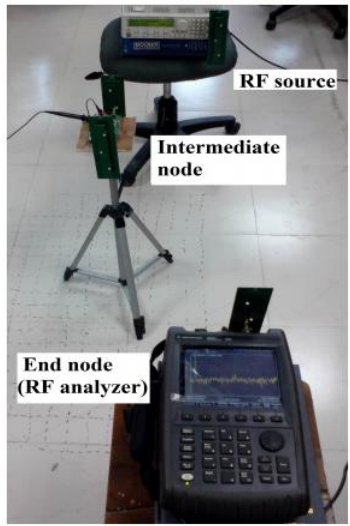

Fig. 4. Signal relaying configuration [11].

Energy trading [12] between sensor nodes can greatly improve the charging efficiency of the system by pluralizing the sources of energy in a WRSN. A charging scheme based on a single mobile charger is likely to cause imbalances in energy intake of the sensor nodes. In a WRSN consisting of dense sensor nodes, overcharged sensor nodes and undercharged sensor nodes can co-exist. Imbalances of sensor nodes can be reduced by energy trading between sensor nodes. Overcharged sensor nodes act as seller nodes while undercharged ones become buyer nodes. The seller nodes charge the buyer nodes, whereas the buyer nodes harvest energy transferred from the seller nodes. With dense and distributed sensor nodes, a buyer node is prone to being surrounded by seller nodes and a seller node is also prone to being surrounded by buyer nodes. Therefore, loss due to off-target charging is reduced. Xiao et al. [12] described an energy trading scheme for sensor nodes to harvest energy from ambience. Each sensor node is aware of its current energy level and how much energy is necessary to perform its duty to transmit messages. The sensor nodes that do not have enough energy to send all their buffered data become buyer nodes, while the sensor nodes that are able to harvest more energy than necessary become seller nodes. A request-and-acceptance process is set up between those two groups of nodes in a one-to-one correspondence. In a WRSN with distributed sensor nodes, this sort of energy trading can be performed in many-to-one correspondence due to many seller nodes around each buyer node.

**C. RF Charging with Mobile Charger**

A typical assumption for charging schemes with mobile chargers is that the mobile charger knows the locations of all the sensor nodes. Charging schemes with mobile chargers must account for path planning for the charger, target energy level of sensor nodes, and possible interaction between sensor nodes. An infrastructure charger is not suitable for use with distributed sensor nodes because the sensor nodes distant from it require a lot more energy to reach the target level compared to the sensor nodes in the charger's immediate vicinity. Figure 5 shows a mobile charger charging distributed sensor nodes in a square service area. Charging scheme should be designed in consideration of the charging efficiency, measured by the ratio of the amount of energy received by the sensor nodes to the amount of energy transmitted by the charger, and the charging time taken for all the sensor nodes to reach individual or common target energy level. Charging efficiency depends on path planning, types of transmit antenna of the charger and receive antenna of the sensor nodes, performance of the RF circuits in the charger and the sensor nodes, polarization loss, and pattern of distributed sensor nodes. The charging time is affected by the factors determining the charging efficiency as well as by the transmit power of the charger.

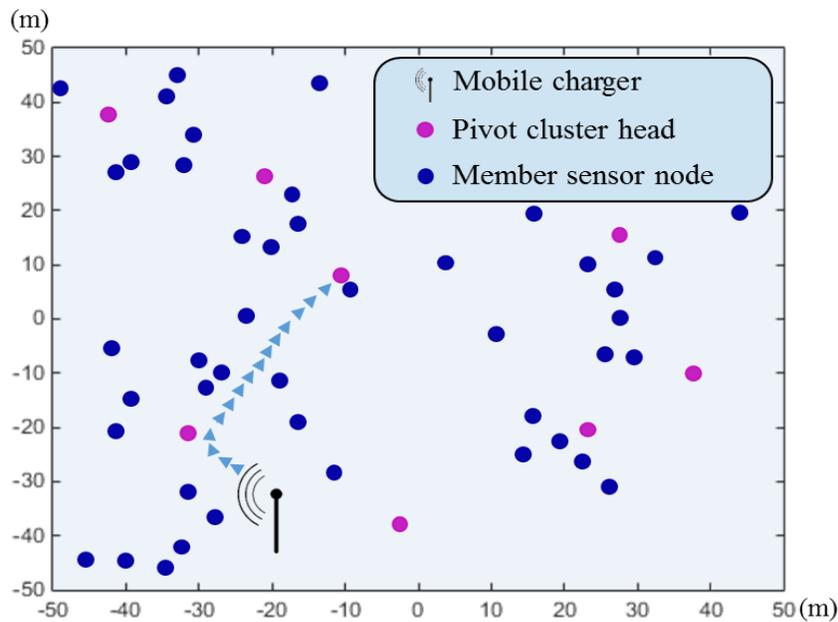

Fig. 5. Mobile charger and distributed sensor nodes.

There are many different ways to provide energy to the sensor nodes in a WRSN. Fu et al. [4] presented a charging scheme with a mobile charger. The goal of their charging scheme is to find optimal locations of the mobile charger that can facilitate minimum charging time for all the sensor nodes to reach the target energy level. Linear programming is used to find the optimal spots in which the charger should be stationed over individual sojourn times. The wireless propagation model, valid for short distances on the order of ten meters, is based on Friss's power equation, as follows:

$$p_r = \frac{G_t G_r \eta}{L_p} \left( \frac{\lambda}{4\pi(d+\beta)} \right)^2 \quad p_0 = \frac{\alpha}{(d+\beta)^2} \quad (1)$$

where $p_r$, $G_t$, $G_r$, $\eta$, $\lambda$, $d$, $\beta$, $L_p$, and $p_0$ are received power, antenna gain of transmitter, antenna gain of receiver, rectifier efficiency, wavelength, distance between transmitter and receiver, a parameter characterizing charging model, polarization loss, and transmit power, respectively. The value of $\alpha$ in eq.(1) is obtained by combining the parameters $G_t$, $G_r$, $\eta$, $\lambda$, $L_p$, and $p_0$. In their charging scheme, the charger is considered to move from one spot to another instantaneously without transition time and the service area is discretized into concentric circle areas. In order to calculate the amount of energy received by each node, each intersection area between the concentric circles is associated with an attenuation factor in relation to each one of the sensor nodes. Madhja et al. [13] proposed a charging scheme employing multiple chargers. Two different classes of mobiles chargers are used. The chargers that belong to the mobile charger class are responsible only for charging the infrastructure sensor nodes; the charger in the super charger class has the duty of charging the other chargers in the mobile charger class as well as the sensor nodes. Wang [14] presented a charging scheme with two different types of energy sources. The sensor nodes are clustered and the cluster heads have the capability of harvesting solar energy while the other nodes are charged by mobile chargers transferring RF energy. The purpose of clustering is to achieve efficient aggregation of data via multi-hop routing from member sensor nodes, so a cluster head is selected from the perspective of minimum hop distance from each member sensor node.

**D. Pivot Cluster Heads and Directional Antenna**

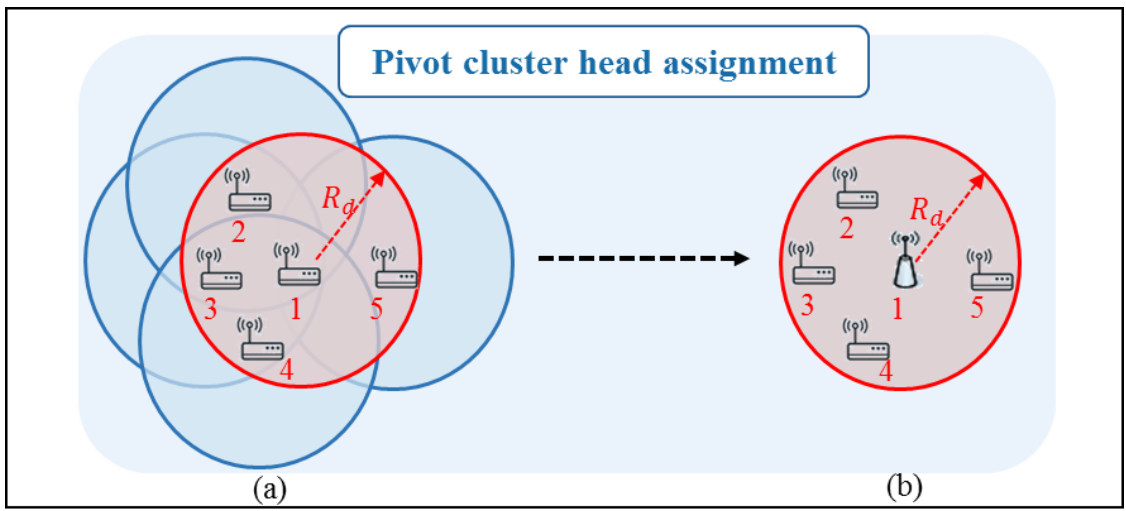

Fig. 6. Clustering process: (a) Selection of pivot cluster heads; (b) pivot cluster head and member sensor nodes.

When sensor nodes are distributed over a service area, a hierarchical charging scheme is more computationally efficient than flat charging because of the smaller number of sensor nodes to be considered for the charging scheme. A hierarchical charging scheme involves multi-stage charging. One possible way to implement a hierarchical charging scheme is to create clusters, each of which consists of a cluster head and member sensor nodes, and charge the cluster heads first. In the next sequence, the charged cluster heads charge the member sensor nodes. As a result, the mobile charger is only concerned with the cluster heads. Member sensor nodes located around the cluster heads are highly likely to be charged as well, due to the radiation pattern of the transmit antenna of the mobile charger. For clustering, widely used clustering algorithms such as the k-means algorithm [15] can be used. However, when the predetermined number of cluster heads is erroneously set, the clustering pattern deviates substantially from visual classification. A simple but efficient clustering process suitable for hierarchical charging is shown in Fig. 6.

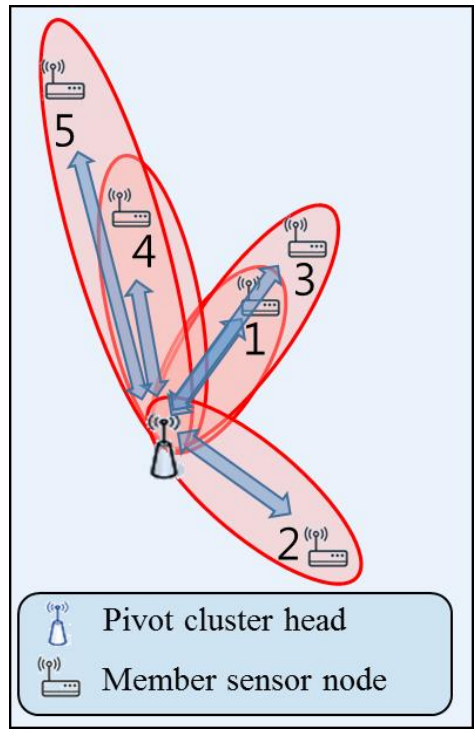

Fig. 7. Sequential charging by pivot cluster head.

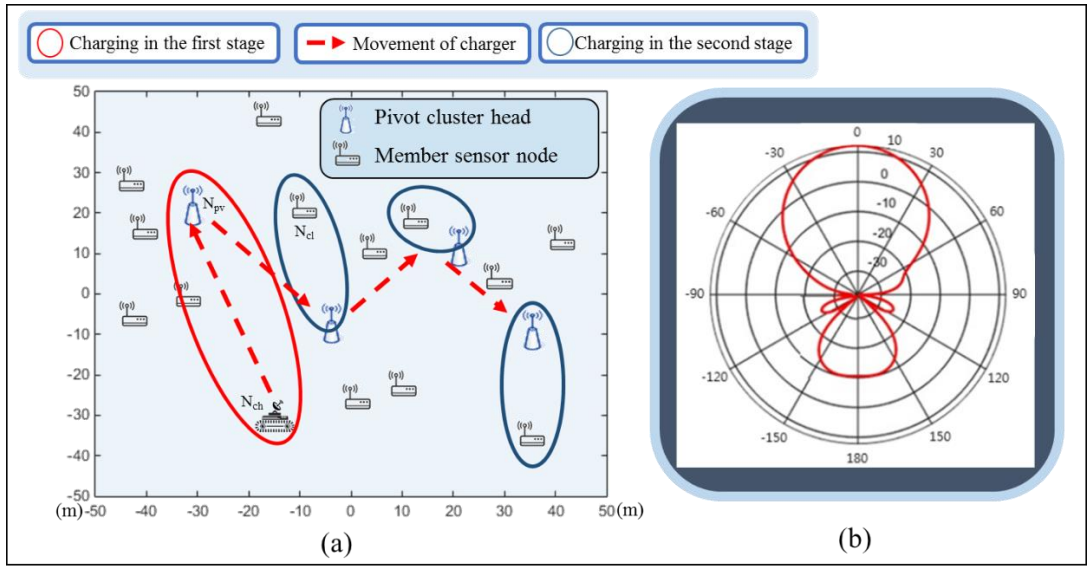

Fig. 8. Pivot cluster heads and directional antenna:(a) distribution of pivot cluster heads and member sensor nodes; (b) directive gain of directional antenna with 12 dB antenna gain.

An inclusion circle is placed with its center at each of 5 sensor nodes to determine the cluster heads. Each sensor node in Fig. 6(a) is associated with an inclusion circle having identical radius $R_d$. The number of sensor nodes within each inclusion circle is counted and the resulting number is compared with the numbers of other circles. The inclusion circle of sensor node 2 includes sensor node 1 and sensor node 3, so the number of sensor nodes within the inclusion circle is 3. Sensor node 1 contains 5 sensor nodes in its inclusion circle and this is the largest number of sensor nodes included by any inclusion circle in the figure. Thus, sensor node 1 is selected as a cluster head. This procedure can be applied to the entire set of sensor nodes in a WRSN. The sensor node associated with the inclusion circle including the largest number of sensor nodes becomes the first cluster head, and the member sensor nodes in the cluster are thus determined. The cluster head and the member sensor nodes in the first cluster are no longer considered for further clustering. To find the second cluster head, the same procedure is repeated with the remaining sensor nodes. This procedure is continued until all the cluster heads are determined. It should be noted that some clusters consist of a single isolated sensor node. Once the locations of cluster heads are determined, directional antennae capable of steering beamforming can be used to make them more powerful chargers in the second stage. When the selected cluster head is able to steer the beamforming with directive gain, the cluster head can charge the nearby sensor nodes sequentially. Figure 7 shows a cluster head charging member sensor nodes sequentially. Such cluster heads are called pivot cluster heads here to emphasize their ability to steer the beamforming. A directional antenna can concentrate the transferred energy in a particular direction or orientation. Let $\Phi_j(t)$ be the orientation angle $\Phi$ of the $j$-th sensor node at time $t$. Then, the wireless propagation model with directive gain $g(\phi_j(t))$ in linear scale is given by:

$$P_r = g\ (\phi_j(t)) \frac{\alpha}{(d_j(t)+\beta)^2} \qquad (2)$$

Eq.(2) can be used for the mobile charger as well as the pivot cluster heads. Figure 8(a) illustrates the two-stage charging process: charging by the mobile charger in the first stage and charging by the pivot cluster heads in the second stage. Figure 8(b) shows the directive gain of the directional antenna according to the orientation angle.

### E. MAC Protocols for Charging by Pivot Cluster Heads

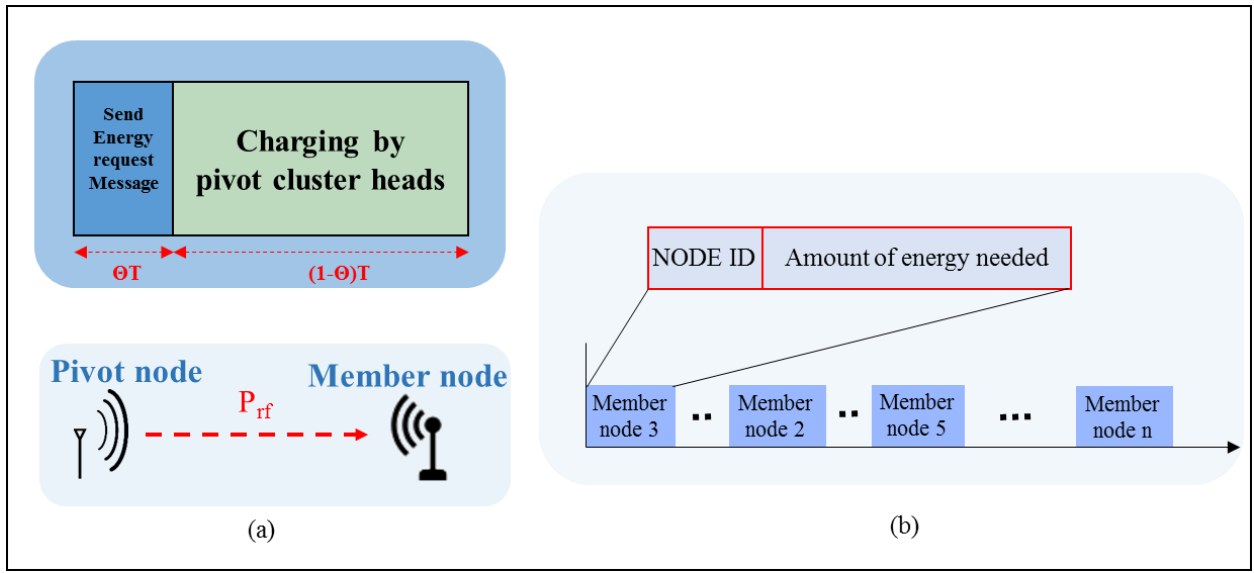

Fig. 9. Charging by pivot cluster heads in the second stage and broadcast messages from undercharged sensor nodes.

A medium access control (MAC) protocol for charging undercharged sensor nodes in the second stage must handle scheduling for broadcast messages from undercharged sensor nodes. Unlike the wired channel, nature of wireless transmission of messages is broadcast, particularly with omni-directional antenna. It is assumed that the frequency band used for broadcast messages is separate from the frequency band for transferring energy. All the undercharged sensor nodes request charging by pivot cluster heads. When the total number of sensor nodes $N_T$ is on the order of hundred, the number of broadcast messages requesting charging is also commensurately large. To handle the flooding of broadcast messages, an efficient MAC protocol should be used. As shown in Fig. 9(a), there is a time interval $\theta T$ for broadcasting the request messages shown in Fig. 9(b). If each pivot cluster head knows the list of member sensor nodes in the same cluster, it can decide the order of charging based on its own policy. The MAC protocol to handle scheduling for broadcast can be polling based, carrier sense multiple access (CSMA) based, or round-robin based. Polling based scheduling [16] based on the initiative taken by the pivot cluster head is not suitable for charging in the second stage, since many undercharged sensor nodes request charging at the same time in the beginning of the second stage. The CSMA based random access [17] of the broadcast channel by the undercharged sensor nodes might be efficient when the number of undercharged sensor nodes is small, but less efficient with a large number of undercharged sensor nodes due to frequent collisions of broadcast messages. Round robin based scheduling [18] can serve well with a large number of undercharged sensor nodes. The order of charging upon receiving requests from sensor nodes can be non-priority based, without respect to the amount of energy needed for each node, or priority-based, serving the sensor nodes with large amount of energy needed ahead of the others.

## III. Energy-efficient Charging in Two Stages with Mobile Charger and Pivot Cluster Heads

### A. Path Planning for Mobile Charger and Charging by Pivot Cluster Heads

A charging scheme in two stages with a mobile charger and pivot cluster heads is described in this section. The total number of sensor nodes is denoted as $N_T$ and the total number of pivot cluster heads is given as $N_p$. Unlike the mobile charger stopped at optimal spots in [4], the mobile charger considered for this charging scheme travels while charging at constant velocity. In order to reduce the charging time of the mobile charger, the clustering algorithm illustrated in Fig. 6 is used to get pivot cluster heads. As a result, each cluster has a pivot cluster head and member sensor nodes. In the first stage, the mobile charger charges the pivot cluster heads and in the second stage the pivot cluster heads charge the member sensor nodes. The pivot cluster heads have one omni-directional receive antenna with 0dB antenna gain and one transmit antenna with 12dB antenna gain, whereas member sensor nodes have both antennae with common 0dB antenna gain. The wireless propagation model for charging the pivot cluster heads, as well as for charging the member sensor nodes, is eq.(2). The directive gain of the transmit antenna is shown in Fig. 8(b). The values of α and β in eq.(2) are set at 36 and 30, respectively, just like those used for [4].

For a given starting spot of the mobile charger, the path of the mobile charger is planned according to the minimized charging time. Let the objective function $T_{cm}(PATH_i)$ be the charging time $T_{cm}$ taken along a path $PATH_i$. Then, the goal of optimization is to minimize the $T_{cm}(PATH_i)$ as follows:

$$T_{cm,\min} = \min_i T_{cm}(PATH_i) \quad \forall paths \quad (3)$$

$$\text{subject to } E_j(T_{cm,min}) \geq ET_p \quad j=1,...,N_p$$

where $E_j(T_{cm,min})$ represents the charged energy level of the $j$-th pivot cluster head and $ET_p$ indicates the target energy level of the pivot cluster heads. Each location of the cluster head is chosen as the starting spot of the mobile charger and for given starting spot the sub-optimal path satisfying smallest charging time is chosen. The optimal path with the locations of cluster heads taken as the respective starting spots is the one requiring minimum charging time among all the sub-optimal paths.

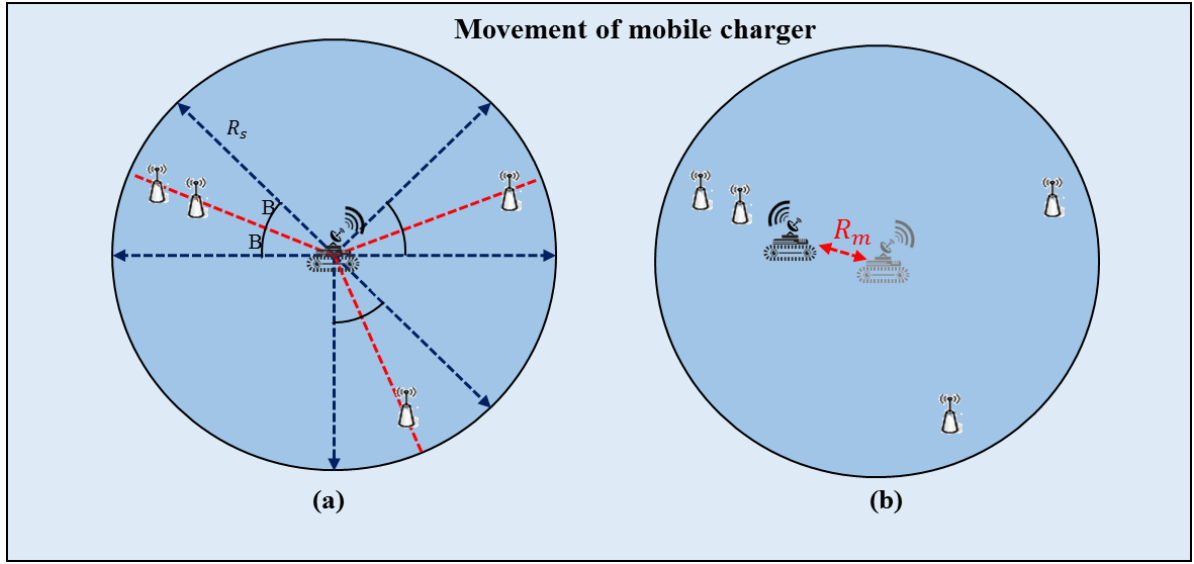

Fig. 10. Movement of mobile charger based on pivot cluster heads.

The charging path of the mobile charger can be modeled by concatenated discrete movements. For each time interval $\Delta t$, a discrete movement action is made by the mobile charger. The direction of the next movement is chosen as the one that yields the largest received power by all the undercharged pivot cluster heads within a sector. The directive gain of the directional antenna can be leveraged to reduce the effort of searching for an optimal direction. When determining the direction (angle) of the next movement of the mobile charger, only the directions toward the undercharged pivot cluster heads are considered. Figure 10(a) indicates potential directions of next movement by dashed lines in red color. The reason why the undercharged cluster heads are considered when making a decision on the next movement is to get them overcharged as quickly as possible to prepare for charging in the second stage. The directive gain of the directional antenna can be leveraged to limit the number of undercharged cluster heads considered for the direction of the charger movement. The interior angle of each service sector is 2B, where B=$22^o$ is the half-power beamwidth of the radiation pattern in Fig. 8(b), and the bisector of the interior angle, corresponding to the potential direction of next movement, connects the charger to an undercharged cluster head. The time interval $\Delta t$ is set at 2.5msec and the distance $R_m$ between the current location and the next location is set at 0.05m. Thus, when the mobile charger keeps shifting, the velocity of the mobile charger is equal to (0.05m/2.5ms)=72km/hour. When staying at the same location yields the largest received power by all the undercharged pivot cluster heads within the same or different sector, the charger decides to do so. Figure 10(b) shows the movement of the mobile charger according to the locations of the pivot cluster heads.

Charging by pivot cluster heads begins when every pivot cluster head becomes overcharged. During the first stage, the mobile charger transmits the RF energy by aiming at the pivot cluster heads, but due to the geophysical proximity of the member sensor nodes to the pivot cluster heads, many member nodes are fairly charged as well. Nevertheless, many member sensor nodes are still undercharged when the first stage is complete. To bring these undercharged member sensor nodes up to the target energy level $ET_s$, the pivot cluster heads charge the member sensor nodes that are still in need of energy. In the beginning of the second stage energy, all the undercharged member sensor nodes broadcast messages, as is shown in Fig. 9(b). It is assumed here that each pivot cluster head knows the list of member sensor nodes in its cluster. After receiving all the messages from the member sensor nodes, each pivot cluster head charges the member sensor nodes in the same cluster. Order of preference in charging can be determined by the amount of energy needed or another criterion. If a member sensor node gets all the energy it needs, it broadcasts a message indicating completion of charging. In the following simulations, MAC protocol to determine the order of charging is not taken into account.

### B. Charging Efficiency of Charging Scheme in Two Stages

The performance of the charging scheme in two stages is evaluated in a given total service area with different random distributions of sensor nodes. The total service area used for the simulations is 100m x100m. This is similar to the area of a factory, a warehouse, or a parking lot, where many wireless sensor nodes are deployed [19][20]. Random distributions of sensor nodes are tested for evaluation of the charging performance and all the sensor nodes are initially energy-depleted. The number of sensor nodes $N_T$ ranges from 100 to 200. The target energy level of the member sensor nodes $ET_s$=2Joules, just like that for [4] and the target energy level of the pivot cluster heads $ET_p$ is adjusted according to $N_T$. The $ET_p$ is set significantly higher than $ET_s$ to guarantee that the pivot cluster heads not only have enough energy for their own tasks but also can charge the member sensor nodes in their vicinity. The charging efficiency $T_f$, the ratio of received power at distance $d=0m$ to transmit power $P_0$ of the pivot cluster head, is set at $T_f =0.02$, as in [22]. The radius of the inclusion circle is set at $R_{cl} =10m$.

Table 1. Simulation parameters

| Types of cluster heads / Charging Configuration | Antenna gain of mobile charger | Transmit antenna gain of pivot cluster heads | $N_T$ | $ET_p$ / $ET_h$ |
|---|---|---|---|---|
| Pivot cluster heads | 12 dB | 12 dB | 200 nodes | 5.5 J |
| | | | 175 nodes | 5.5 J |
| | | | 150 nodes | 4.5 J |
| | | | 125 nodes | 4.5 J |
| | | | 100 nodes | 4.0 J |
| Cluster heads | 12 dB | 0 dB | 200 nodes | 5.0 J |
| | | | 175 nodes | 5.5 J |
| | | | 150 nodes | 5.5 J |
| | | | 125 nodes | 6.0 J |
| | | | 100 nodes | 6.0 J |

The performance of the charging scheme in two stages is compared with that of other charging schemes. One of the charging schemes for comparison is the scheme in [4]. Another charging scheme is also in two stages with a difference in that there is an absence of pivot cluster heads with directional antennae. Path planning of the mobile charger in the first stage is designed to charge cluster heads. Antenna gain of the transmit antennae of the cluster heads is 0dB instead of 12dB for the pivot cluster heads. When charging with the mobile charger in the first stage is finished, every cluster head becomes overcharged. Unlike charging only by pivot cluster heads in the second stage, all the overcharged sensor nodes, including the cluster heads, can charge the undercharged member sensor nodes. The target energy level $ET_h$ with the overcharged sensor nodes in the second stage is also set according to $N_T$, as shown in Table 1. Since the cluster heads employ transmit antennae with 0 dB antenna gain, eq.(1) is used instead of eq.(2). From Table 1, it is seen that the $ET_p$ is set higher than the $ET_h$ with high density of sensor nodes, but it is set lower with low density of sensor nodes. This is because with larger $N_T$ the number of overcharged sensor nodes increases, so that there are more sensor nodes that are able to charge undercharged sensor nodes.

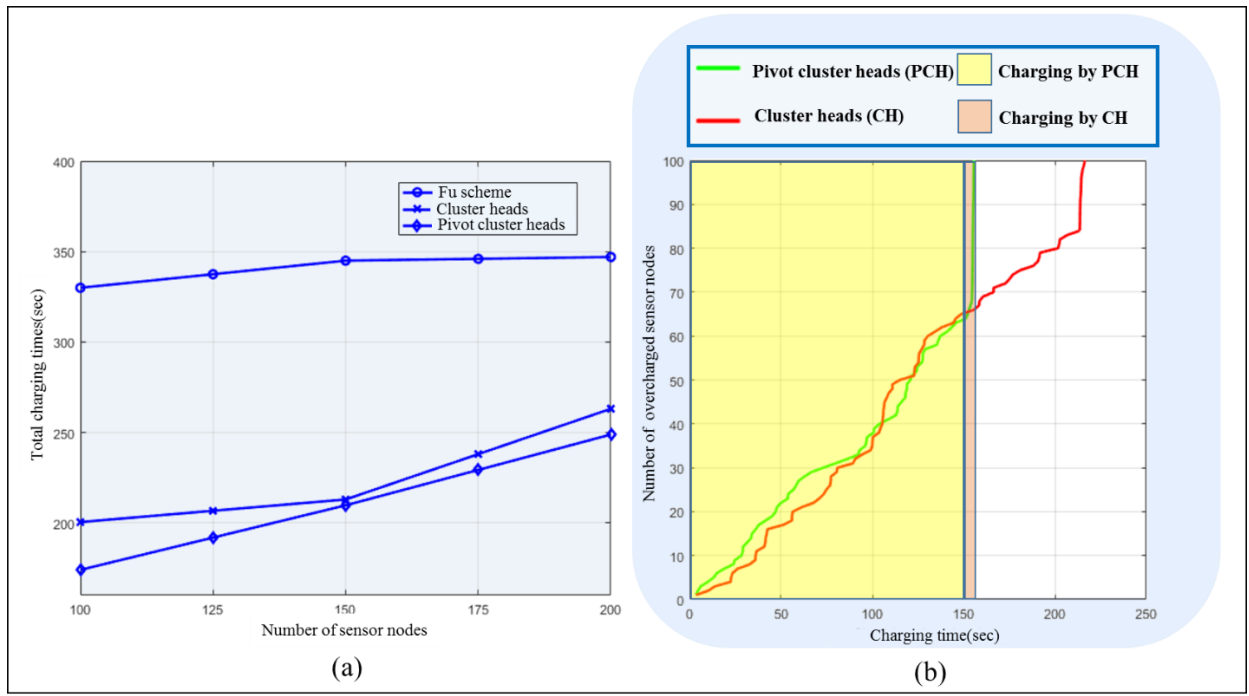

Fig.11. Simulation results: (a) total charging time; (b) number of overcharged sensor nodes as a function of time

Figure 11 shows the total charging time for the three schemes considered. Both charging schemes in two stages outperform the charging scheme in [4] and the charging scheme involving pivot cluster heads provides the best performance. Since the pivot cluster heads use directional antennae, the charging efficiency in the second stage is significantly higher than that of the other schemes in two stages. On the other hand, charging efficiency with cluster heads of omni-directional antennae in the second stage is seen comparable to that with pivot cluster heads employing directional antennae. Compared with the scheme in [4], charging by overcharged sensor nodes in the second stage seems effective at reducing the total charging time. Figure 11(b) shows the charging progress in terms of the number of charged sensor nodes over time. Figure 11(b) suggests that the charging scheme with overcharged sensor nodes for charging in the second stage requires substantially more sensor nodes to be overcharged in the first stage as compared to the case with the pivot cluster heads. This is because of the waste of energy that occurs when the overcharged sensor nodes transfer energy and the waste occurs due to the omni-directional antenna incurring much transmission loss along wrong directions. Thus, the number of undercharged sensor nodes that can become overcharged in the second stage is substantially smaller with the overcharged sensor nodes used in the second stage. The simulation results obtained here are in accordance with the results in [9], where it is stated that a directional antenna is more efficient than an omni-directional antenna for charging applications. Figure 12 shows the energy profile of the pivot cluster heads before and after the second stage with three different $N_T$ values. The amount of surplus energy of a pivot cluster head increases with increased $N_T$, and the surplus energy becomes approximately 0 after the second stage.

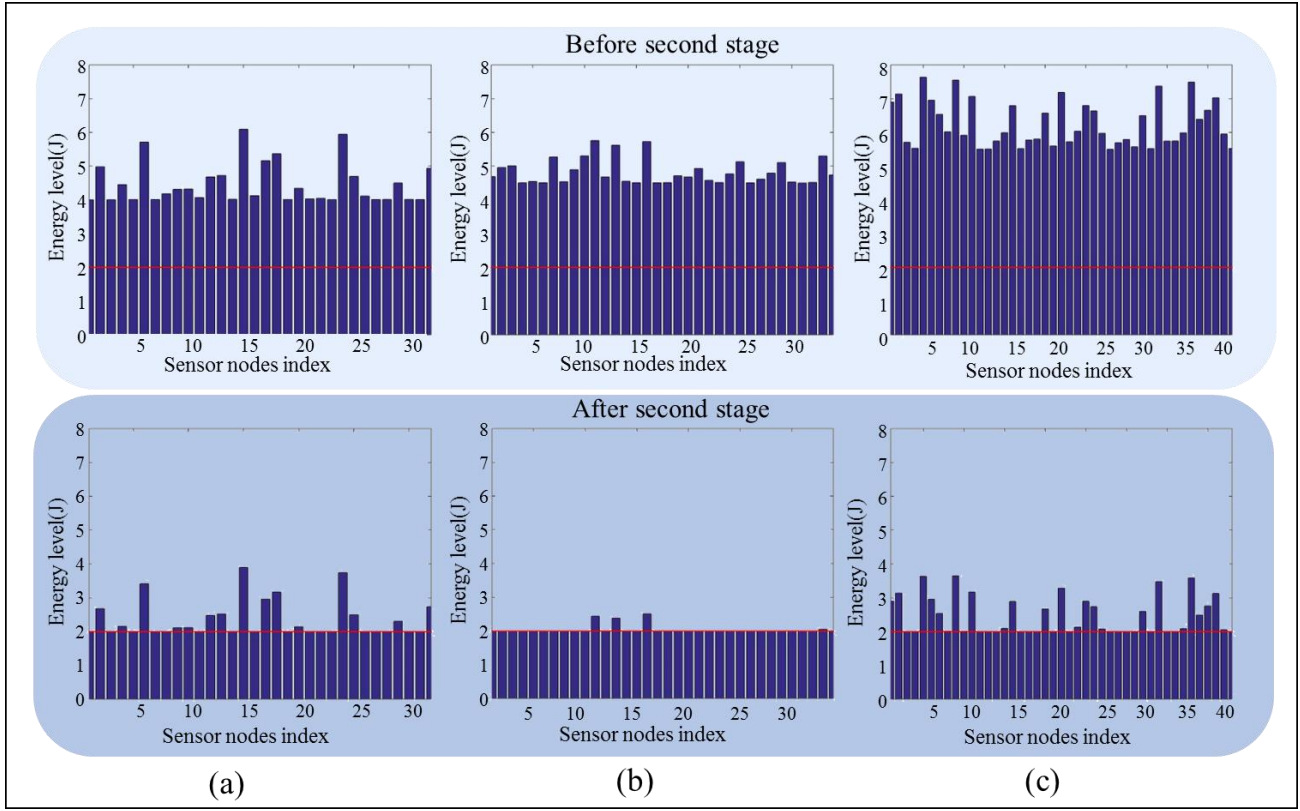

Fig.12. Energy level of pivot cluster heads before and after the second stage: (a) $N_T$=100; (b) $N_T$=150; (c) $N_T$=200.

## IV. CONCLUSION

Review of wireless rechargeable sensor networks (WRSNs) is presented in this article. To ensure sustainable operation of sensor networks, charging efficiency is considered important. With the large number of existing WRSNs, energy-efficient charging schemes become critical to workplaces aiming at low operating cost. System aspects of RF remote charging and RF energy harvesting are described. To obtain higher charging efficiency, a directional antenna providing higher antenna gain can be adopted.

The RF remote charging scheme consisting of charging pivot cluster heads by a mobile charger with directional antenna in the first stage and charging undercharged sensor nodes by pivot cluster heads with directional antenna in the second stage is presented. Also, another charging scheme involving energy trading between overcharged sensor nodes and undercharged sensor nodes in the second stage is presented. The proposed charging schemes are expected to be particularly useful in open spaces with many sensor nodes such as factories, warehouses, and parking lots, where sensor nodes can be densely deployed.